\pdfoutput=1

\documentclass[12pt,a4paper]{article}

\usepackage{ifthen} 
\newboolean{pdflatex}
\setboolean{pdflatex}{true} 

\newboolean{articletitles}
\setboolean{articletitles}{true} 

\newboolean{uprightparticles}
\setboolean{uprightparticles}{false} 

\usepackage{microtype}
\usepackage{lineno}  
\usepackage{xspace} 

\usepackage{graphicx}  
\usepackage{color}
\usepackage{colortbl}
\graphicspath{{./figs/}} 

\usepackage{amsmath} 
\usepackage{amssymb}
\usepackage{amsfonts}
\usepackage{upgreek} 

\newcommand*\patchAmsMathEnvironmentForLineno[1]{%
\expandafter\let\csname old#1\expandafter\endcsname\csname #1\endcsname
\expandafter\let\csname oldend#1\expandafter\endcsname\csname
end#1\endcsname
 \renewenvironment{#1}%
   {\linenomath\csname old#1\endcsname}%
   {\csname oldend#1\endcsname\endlinenomath}%
}
\newcommand*\patchBothAmsMathEnvironmentsForLineno[1]{%
  \patchAmsMathEnvironmentForLineno{#1}%
  \patchAmsMathEnvironmentForLineno{#1*}%
}
\AtBeginDocument{%
\patchBothAmsMathEnvironmentsForLineno{equation}%
\patchBothAmsMathEnvironmentsForLineno{align}%
\patchBothAmsMathEnvironmentsForLineno{flalign}%
\patchBothAmsMathEnvironmentsForLineno{alignat}%
\patchBothAmsMathEnvironmentsForLineno{gather}%
\patchBothAmsMathEnvironmentsForLineno{multline}%
}

\usepackage{hyperref}    
\usepackage[all]{hypcap} 

\def\lhcb {\mbox{LHCb}\xspace}
\def\ux85 {\mbox{UX85}\xspace}

\def\babar  {\mbox{BaBar}\xspace}
\def\belle  {\mbox{Belle}\xspace}

\ifthenelse{\boolean{uprightparticles}}%
{
 
 \def\Pgamma      {\ensuremath{\upgamma}\xspace}

 \def\Pmu         {\ensuremath{\upmu}\xspace}

 \def\Ppsi        {\ensuremath{\uppsi}\xspace}

 \def\PDelta      {\ensuremath{\Delta}\xspace}                 
 \def\PXi      {\ensuremath{\Xi}\xspace}                 
 \def\PLambda      {\ensuremath{\Lambda}\xspace}                 
 \def\PSigma      {\ensuremath{\Sigma}\xspace}                 
 \def\POmega      {\ensuremath{\Omega}\xspace}                 
 \def\PUpsilon      {\ensuremath{\Upsilon}\xspace}                 
 
 \def\PB      {\ensuremath{\mathrm{B}}\xspace}                 
                  
 \def\PD      {\ensuremath{\mathrm{D}}\xspace}

 \def\PJ      {\ensuremath{\mathrm{J}}\xspace}                 
 \def\PK      {\ensuremath{\mathrm{K}}\xspace}

 \def\Pb      {\ensuremath{\mathrm{b}}\xspace}                 
 \def\Pc      {\ensuremath{\mathrm{c}}\xspace}                 
                  
 \def\Pe      {\ensuremath{\mathrm{e}}\xspace}

 \def\Pi      {\ensuremath{\mathrm{i}}\xspace}

 \def\Ps      {\ensuremath{\mathrm{s}}\xspace}

}
{
 
 \def\Pgamma      {\ensuremath{\gamma}\xspace}

 \def\Pmu         {\ensuremath{\mu}\xspace}

 \def\Ppsi        {\ensuremath{\psi}\xspace}                 
                  
 \mathchardef\PDelta="7101
 \mathchardef\PXi="7104
 \mathchardef\PLambda="7103
 \mathchardef\PSigma="7106
 \mathchardef\POmega="710A
 \mathchardef\PUpsilon="7107
                  
 \def\PB      {\ensuremath{B}\xspace}                 
                  
 \def\PD      {\ensuremath{D}\xspace}

 \def\PJ      {\ensuremath{J}\xspace}                 
 \def\PK      {\ensuremath{K}\xspace}

 \def\Pb      {\ensuremath{b}\xspace}                 
 \def\Pc      {\ensuremath{c}\xspace}                 
                  
 \def\Pe      {\ensuremath{e}\xspace}

 \def\Pi      {\ensuremath{i}\xspace}

 \def\Ps      {\ensuremath{s}\xspace}

}

\def\en         {\ensuremath{\Pe^-}\xspace}   
\def\ep         {\ensuremath{\Pe^+}\xspace}
 
\def\epem       {\ensuremath{\Pe^+\Pe^-}\xspace}

\def\mup        {\ensuremath{\Pmu^+}\xspace}
\def\mun        {\ensuremath{\Pmu^-}\xspace} 

\def\ellm       {\ensuremath{\ell^-}\xspace}
\def\ellp       {\ensuremath{\ell^+}\xspace}

\def\g      {\ensuremath{\Pgamma}\xspace}

\def\squark    {\ensuremath{\Ps}\xspace}

\def\cquark    {\ensuremath{\Pc}\xspace}

\def\bquark    {\ensuremath{\Pb}\xspace}

\def\kaon  {\ensuremath{\PK}\xspace}
  \def\Kbar  {\kern 0.2em\overline{\kern -0.2em \PK}{}\xspace}

\def\Kz    {\ensuremath{\kaon^0}\xspace}
\def\Kzb   {\ensuremath{\Kbar^0}\xspace}
\def\KzKzb {\ensuremath{\Kz \kern -0.16em \Kzb}\xspace}
\def\Kp    {\ensuremath{\kaon^+}\xspace}
\def\Km    {\ensuremath{\kaon^-}\xspace}

\def\KpKm  {\ensuremath{\Kp \kern -0.16em \Km}\xspace}

\def\Kstarz  {\ensuremath{\kaon^{*0}}\xspace}

\def\Kstar   {\ensuremath{\kaon^*}\xspace}

  \def\Dbar    {\kern 0.2em\overline{\kern -0.2em \PD}{}\xspace}
\def\D       {\ensuremath{\PD}\xspace}

\def\Dz      {\ensuremath{\D^0}\xspace}
\def\Dzb     {\ensuremath{\Dbar^0}\xspace}
\def\DzDzb   {\ensuremath{\Dz {\kern -0.16em \Dzb}}\xspace}
\def\Dp      {\ensuremath{\D^+}\xspace}
\def\Dm      {\ensuremath{\D^-}\xspace}

\def\DpDm    {\ensuremath{\Dp {\kern -0.16em \Dm}}\xspace}

\def\Dstarp  {\ensuremath{\D^{*+}}\xspace}

\def\B       {\ensuremath{\PB}\xspace}
\def\Bbar    {\ensuremath{\kern 0.18em\overline{\kern -0.18em \PB}{}}\xspace}

\def\Bz      {\ensuremath{\B^0}\xspace}

\def\Bd      {\ensuremath{\B^0}\xspace}
\def\Bs      {\ensuremath{\B^0_\squark}\xspace}

\def\jpsi     {\ensuremath{{\PJ\mskip -3mu/\mskip -2mu\Ppsi\mskip 2mu}}\xspace}
\def\psitwos  {\ensuremath{\Ppsi{(2S)}}\xspace}

  \def\Y#1S{\ensuremath{\PUpsilon{(#1S)}}\xspace}

\def\L {\ensuremath{\PLambda}\xspace}
\def\Lbar {\ensuremath{\kern 0.1em\overline{\kern -0.1em\PLambda}}\xspace}

\def\Lb      {\ensuremath{\L^0_\bquark}\xspace}

\def\BF         {{\ensuremath{\cal B}\xspace}}

\def\BR         {\BF}
\newcommand{\decay}[2]{\ensuremath{#1\!\to #2}\xspace}         
\def\ra                 {\ensuremath{\rightarrow}\xspace}
\def\to                 {\ensuremath{\rightarrow}\xspace}

\newcommand{\mBd}{\ensuremath{m_{\Bd}}\xspace}

\def\qsq       {\ensuremath{q^2}\xspace}

\def\BdToKstmm    {\decay{\Bd}{\Kstarz\mup\mun}}

\def\BdToJPsiKst  {\decay{\Bd}{\jpsi\Kstarz}}

\def\BdKstGam     {\decay{\Bd}{\Kstarz \g}}

\def\BdKstee  {\decay{\Bd}{\Kstarz\epem}}
\def\BdToJPsieeKst  {\decay{\Bd}{\jpsi(\epem)\Kstarz}}

\def\BsPhiee  {\decay{\Bs}{\phi\epem}}

\def\BsToJPsieePhi  {\decay{\Bs}{\jpsi(\epem)\phi}}

\def\JPsiToee  {\decay{\jpsi}{\epem}}

\def\BdKstll     {\decay{\Bd}{\Kstarz \ell^+ \ell^-}}

\def\AT#1     {\ensuremath{A_{\mathrm{T}}^{#1}}\xspace}           
\def\btosgam  {\decay{\bquark}{\squark \g}}

\def\C#1      {\ensuremath{\mathcal{C}_{#1}}\xspace}                       
\def\Cp#1     {\ensuremath{\mathcal{C}_{#1}^{'}}\xspace}                    
\def\Ceff#1   {\ensuremath{\mathcal{C}_{#1}^{\mathrm{(eff)}}}\xspace}        
\def\Cpeff#1  {\ensuremath{\mathcal{C}_{#1}^{'\mathrm{(eff)}}}\xspace}       
\def\Ope#1    {\ensuremath{\mathcal{O}_{#1}}\xspace}                       
\def\Opep#1   {\ensuremath{\mathcal{O}_{#1}^{'}}\xspace}                    

\newcommand{\tev}{\ensuremath{\mathrm{\,Te\kern -0.1em V}}\xspace}
\newcommand{\gev}{\ensuremath{\mathrm{\,Ge\kern -0.1em V}}\xspace}
\newcommand{\mev}{\ensuremath{\mathrm{\,Me\kern -0.1em V}}\xspace}
\newcommand{\kev}{\ensuremath{\mathrm{\,ke\kern -0.1em V}}\xspace}
\newcommand{\ev}{\ensuremath{\mathrm{\,e\kern -0.1em V}}\xspace}
\newcommand{\gevc}{\ensuremath{{\mathrm{\,Ge\kern -0.1em V\!/}c}}\xspace}
\newcommand{\mevc}{\ensuremath{{\mathrm{\,Me\kern -0.1em V\!/}c}}\xspace}
\newcommand{\gevcc}{\ensuremath{{\mathrm{\,Ge\kern -0.1em V\!/}c^2}}\xspace}
\newcommand{\gevgevcccc}{\ensuremath{{\mathrm{\,Ge\kern -0.1em V^2\!/}c^4}}\xspace}
\newcommand{\mevcc}{\ensuremath{{\mathrm{\,Me\kern -0.1em V\!/}c^2}}\xspace}

\def\mum  {\ensuremath{\,\upmu\rm m}\xspace}

\def\invfb   {\ensuremath{\mbox{\,fb}^{-1}}\xspace}

\newcommand{\chisq}{\ensuremath{\chi^2}\xspace}

\def\gsim{{~\raise.15em\hbox{$>$}\kern-.85em
          \lower.35em\hbox{$\sim$}~}\xspace}
\def\lsim{{~\raise.15em\hbox{$<$}\kern-.85em
          \lower.35em\hbox{$\sim$}~}\xspace}

\def\sPlot{\mbox{\em sPlot}}

\def\pt         {\mbox{$p_{\rm T}$}\xspace}

\def\evtgen     {\mbox{\textsc{EvtGen}}\xspace}
\def\pythia     {\mbox{\textsc{Pythia}}\xspace}

\def\geant      {\mbox{\textsc{Geant4}}\xspace}

\def\photos     {\mbox{\textsc{Photos}}\xspace}

\def\tell1  {TELL1\xspace}
\def\ukl1   {UKL1\xspace}

\usepackage{cite} 
\usepackage{mciteplus}

\usepackage{longtable} 

\begin{document}

\renewcommand{\thefootnote}{\fnsymbol{footnote}}
\setcounter{footnote}{1}



\begin{titlepage}
\pagenumbering{roman}

\vspace*{-1.5cm}
\centerline{\large EUROPEAN ORGANIZATION FOR NUCLEAR RESEARCH (CERN)}
\vspace*{1.5cm}
\hspace*{-0.5cm}
\begin{tabular*}{\linewidth}{lc@{\extracolsep{\fill}}r}
\ifthenelse{\boolean{pdflatex}}
{\vspace*{-2.7cm}\mbox{\!\!\!\includegraphics[width=.14\textwidth]{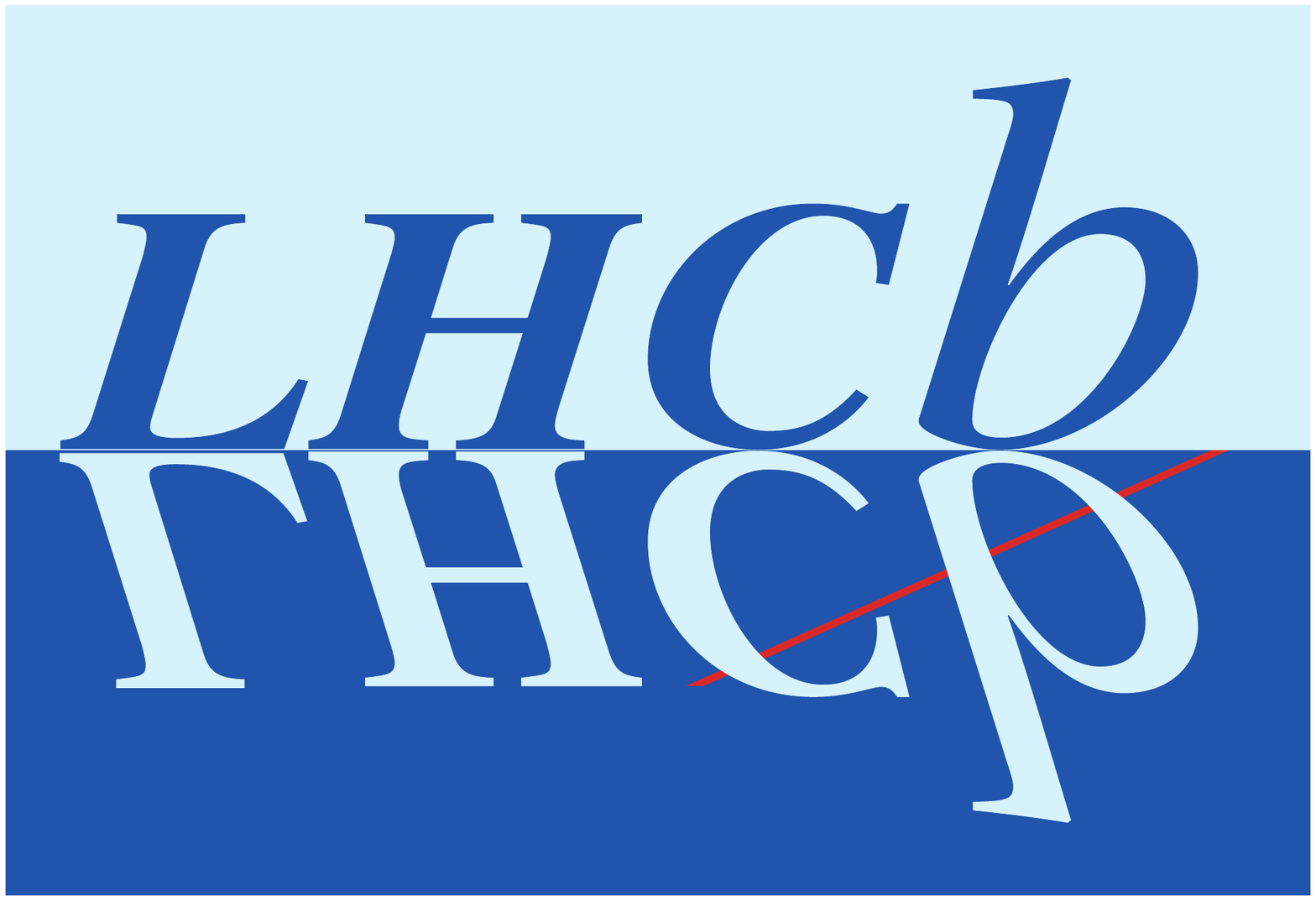}} & &}%
{\vspace*{-1.2cm}\mbox{\!\!\!\includegraphics[width=.12\textwidth]{lhcb-logo.eps}} & &}%
\\
 & & CERN-PH-EP-2013-048 \\  
 & & LHCb-PAPER-2013-005 \\  
 & & April 10, 2013 \\
 & & \\
\end{tabular*}

\vspace*{4.0cm}

{\bf\boldmath\huge
\begin{center}
  Measurement of the \BdKstee  branching fraction at low dilepton mass
  
\end{center}
}

\vspace*{1.0cm}

\begin{center}
The LHCb collaboration\footnote{Authors are listed on the following pages.}
\end{center}

\vspace{\fill}

\begin{abstract}
  \noindent
 The branching fraction of the rare decay \BdKstee in the dilepton mass region from 30 to 1000 \mevcc has been measured by the \lhcb experiment, using $pp$ collision data, corresponding to an integrated luminosity of 1.0~\invfb, at a centre-of-mass energy of 7 \tev. 
 The decay mode $\Bd \ra \jpsi(\epem) \Kstarz $  is utilized as a normalization channel. The branching fraction ${ \BF(\Bd \ra \Kstarz \epem})$ is measured to be
$$
\BF(\BdKstee)^{30-1000 \mevcc}= (3.1\, ^{+0.9\mbox{ } +0.2}_{-0.8\mbox{ }-0.3 }  \pm 0.2)\times 10^{-7},
$$
where the first error is statistical, the second is systematic, and the third comes from the uncertainties on the \BdToJPsiKst and \JPsiToee  branching fractions.
\end{abstract}
\vspace*{1.0cm}

\begin{center}
  Submitted to JHEP
\end{center}

\vspace{\fill}

{\footnotesize 
\centerline{\copyright~CERN on behalf of the \lhcb collaboration, license \href{http://creativecommons.org/licenses/by/3.0/}{CC-BY-3.0}.}}
\vspace*{2mm}

\end{titlepage}


\newpage
\setcounter{page}{2}
\mbox{~}
\newpage

\centerline{\large\bf LHCb collaboration}
\begin{flushleft}
\small
R.~Aaij$^{40}$, 
C.~Abellan~Beteta$^{35,n}$, 
B.~Adeva$^{36}$, 
M.~Adinolfi$^{45}$, 
C.~Adrover$^{6}$, 
A.~Affolder$^{51}$, 
Z.~Ajaltouni$^{5}$, 
J.~Albrecht$^{9}$, 
F.~Alessio$^{37}$, 
M.~Alexander$^{50}$
% ...

{\footnotesize \it
% ...
$ ^{5}$Clermont Universit\'{e}, Universit\'{e} Blaise Pascal, CNRS/IN2P3, LPC, Clermont-Ferrand, France\\
$ ^{6}$CPPM, Aix-Marseille Universit\'{e}, CNRS/IN2P3, Marseille, France\\
% ...
$ ^{9}$Fakult\"{a}t Physik, Technische Universit\"{a}t Dortmund, Dortmund, Germany\\
% ...
$ ^{35}$Universitat de Barcelona, Barcelona, Spain\\
$ ^{36}$Universidad de Santiago de Compostela, Santiago de Compostela, Spain\\
$ ^{37}$European Organization for Nuclear Research (CERN), Geneva, Switzerland\\
% ...
$ ^{40}$Nikhef National Institute for Subatomic Physics, Amsterdam, The Netherlands\\
% ...
$ ^{45}$H.H. Wills Physics Laboratory, University of Bristol, Bristol, United Kingdom\\
% ...
$ ^{50}$School of Physics and Astronomy, University of Glasgow, Glasgow, United Kingdom\\
$ ^{51}$Oliver Lodge Laboratory, University of Liverpool, Liverpool, United Kingdom\\
% ...
$ ^{n}$LIFAELS, La Salle, Universitat Ramon Llull, Barcelona, Spain\\
% ...
}
\end{flushleft}

\cleardoublepage


\renewcommand{\thefootnote}{\arabic{footnote}}
\setcounter{footnote}{0}



\pagestyle{plain} 
\setcounter{page}{1}
\pagenumbering{arabic}


\section{Introduction}
\label{sec:Intro}
\begin{figure}[b]
\includegraphics[width=0.5\textwidth]{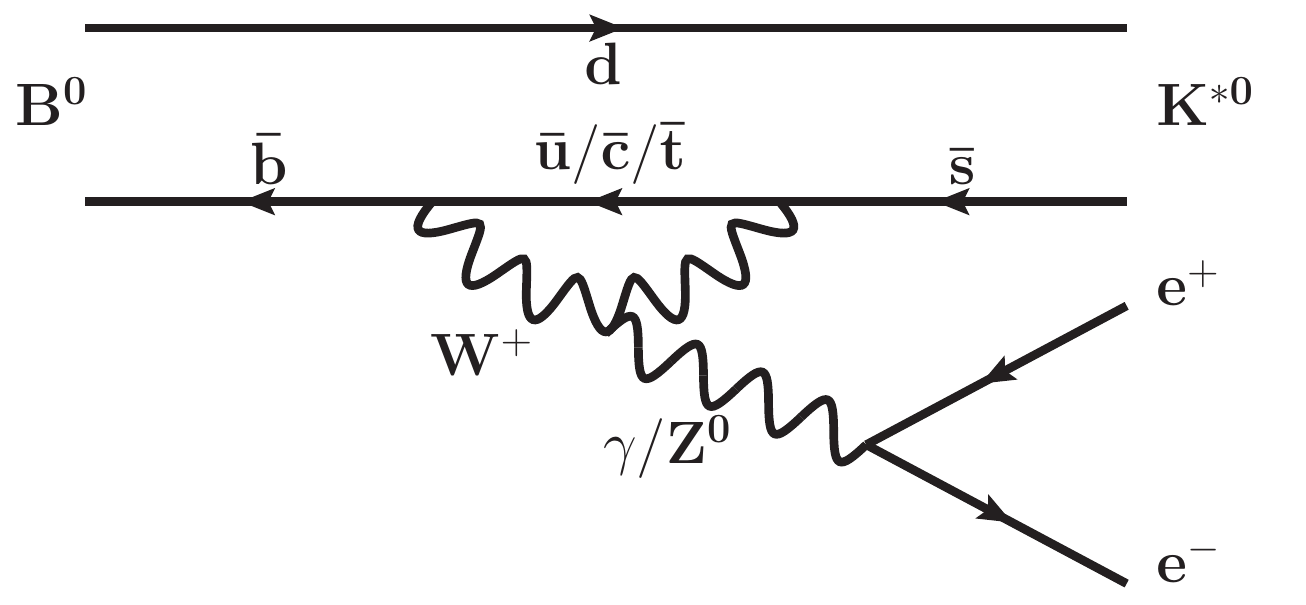}
\includegraphics[width=0.5\textwidth]{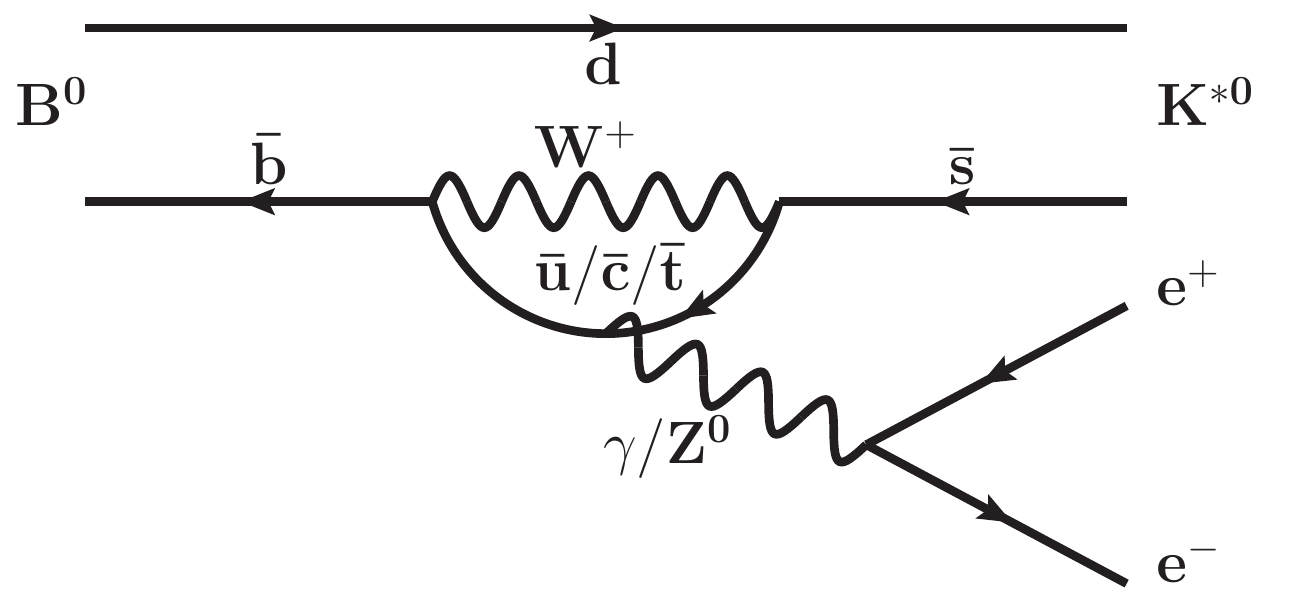}
\vskip 5mm
\includegraphics[width=0.5\textwidth]{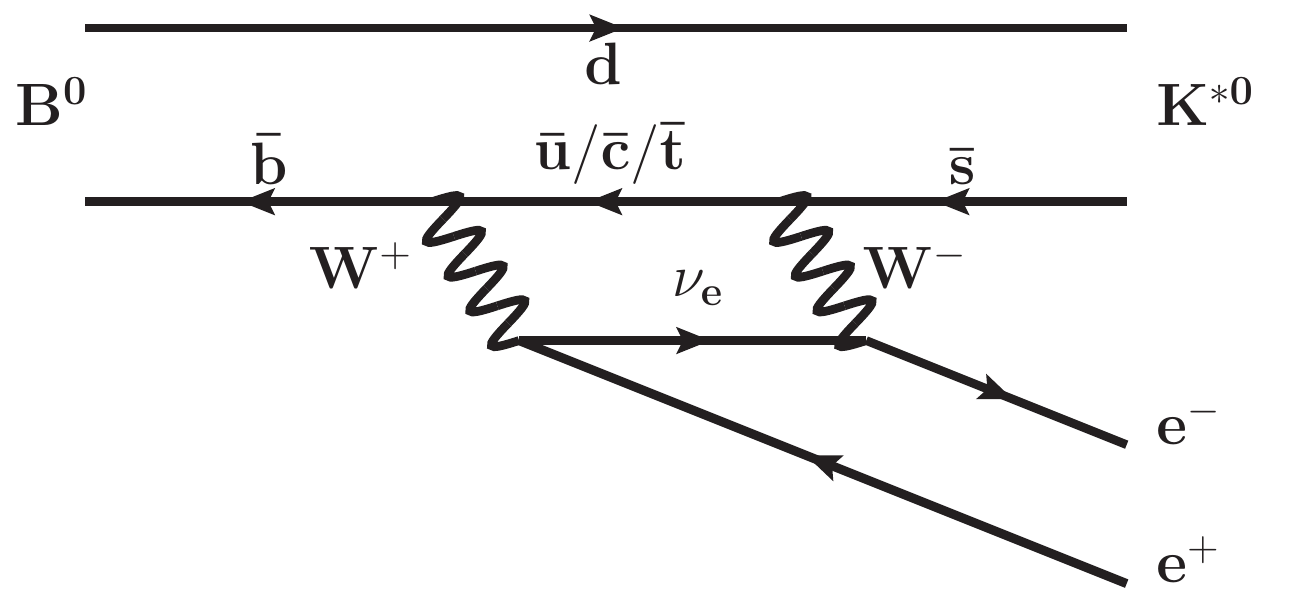}
  \caption{\small Dominant Standard Model diagrams contributing to the decay ${\Bd \ra \Kstarz \epem}$.}
  \label{fig:feynman}
  \end{figure}

The \btosgam transition proceeds through flavour changing neutral currents, and thus is sensitive to the effects of physics beyond the Standard Model (BSM). Although the branching fraction of the \BdKstGam decay has been measured~\cite{Aubert:2009ak,Nakao:2004th,Coan:1999kh} to be consistent with the Standard Model (SM) prediction~\cite{Ali:2007sj}, BSM effects could still be present and detectable through more detailed studies of the decay process. In particular, in the SM the photon helicity is predominantly left-handed, with a small right-handed current arising from long distance effects and from the non-zero value of the ratio of the \squark-quark mass to the \bquark-quark mass. Information on the photon polarisation can be obtained with an angular analysis of the \BdKstll decay ($\ell = e, \mu$) in the low dilepton invariant mass squared (\qsq) region where the photon contribution dominates. The inclusion of charge-conjugate modes is implied throughout the paper. The low \qsq region also has the benefit of reduced theoretical uncertainties due to long distance contributions compared to the full \qsq region~\cite{Grossman:2000rk}. The more precise SM prediction allows for increased sensitivity to contributions from BSM. In the low \qsq  interval there is a contribution from $\Bz \ra  \Kstarz V (V\ra \ell^+\ell^-)$ where $V$ is one of the vector
resonances $\rho$, $\omega$ or $\phi$; however this contribution has been calculated to be at most  1\%~\cite{Korchin:2010uc}. The diagrams contributing to the \BdKstee decay are shown in  Fig. \ref{fig:feynman}. 

With the \lhcb detector, the \BdKstll analysis can be carried out using either muons~\cite{LHCb-PAPER-2011-020} or electrons. Experimentally, the decay with muons in the final state produces a much higher yield per unit integrated luminosity than electrons, primarily due to the clean trigger signature. In addition, the much smaller bremsstrahlung radiation leads to better momentum resolution, allowing a more efficient selection. 
On the other hand, the \BdKstee decay probes lower dilepton invariant masses, thus providing greater sensitivity to the photon 
polarisation~\cite{Grossman:2000rk}. Furthermore, the formalism is greatly simplified due to the negligible lepton mass~\cite{Kruger:2005ep}. It is therefore interesting to carry out an angular analysis of the decay \BdKstee in the region where the dilepton mass is less than 1000\mevcc.  The lower limit is set to 30\mevcc since below this value the sensitivity for the angular analysis decreases because of a degradation in the precision of the orientation of the \epem decay plane due to multiple scattering. Furthermore, the contamination from the \BdKstGam decay, with the photon converting into an \epem pair in the detector material, increases significantly as $\qsq \to 0$. 

The first step towards performing the angular analysis is to measure the branching fraction in this very low dilepton invariant mass region. Indeed, even if there is no doubt about the existence of this decay, no clear \BdKstee signal  has been observed in this region and therefore the partial branching fraction is unknown.  The only experiments to have observed \BdKstee to date are \babar \cite{:2012vwa} and \belle \cite{Wei:2009zv}, which have collected about 30 \decay{\Bd}{\Kstarz\ellp\ellm} events each in the region $\qsq < 2 \gevgevcccc$, summing over electron and muon final states.

\section{The \lhcb detector, dataset and analysis strategy}
\label{sec:Detector}
The study reported here is based on $pp$ collision data, corresponding to an integrated luminosity of 1.0~\invfb, collected at the Large Hadron Collider 
(LHC) with the \lhcb detector~\cite{Alves:2008zz} at a centre-of-mass energy of 7 TeV during 2011.
The \lhcb detector is a single-arm forward
spectrometer covering the \mbox{pseudorapidity} range $2<\eta <5$,
designed for the study of particles containing \bquark or \cquark
quarks. It includes a high precision tracking system
consisting of a silicon-strip vertex detector (VELO) surrounding the $pp$
interaction region, a large-area silicon-strip detector located
upstream of a dipole magnet with a bending power of about
$4{\rm\,Tm}$, and three stations of silicon-strip detectors and straw
drift tubes placed downstream. The combined tracking system has 
momentum resolution $(\Delta p/p)$ that varies from 0.4\% at 5\gevc to
0.6\% at 100\gevc, and impact parameter (IP) resolution of 20\mum for
tracks with high transverse momentum (\pt). Charged hadrons are identified
using two ring-imaging Cherenkov detectors. Photon, electron and
hadron candidates are identified by a calorimeter system consisting of
scintillating-pad (SPD) and preshower (PS) detectors, an electromagnetic
calorimeter (ECAL) and a hadronic calorimeter. Muons are identified by a
system composed of alternating layers of iron and multiwire
proportional chambers. The trigger~\cite{LHCb-DP-2012-004} consists of a
hardware stage, based on information from the calorimeter and muon
systems, followed by a software stage which applies a full event
reconstruction.

For signal candidates to be considered in this analysis, at least one of the electrons from the \BdKstee decay must
pass the hardware electron trigger, or the hardware trigger must be satisfied independently of any of the daughters
of the signal \Bz candidate (usually triggering on the other \bquark-hadron in the event). The hardware electron trigger requires the presence of an ECAL 
cluster with a transverse energy greater than 2.5 GeV. An energy deposit is also required in
one of the PS cells in front of the ECAL cluster, where the threshold
corresponds to the energy that would be deposited by the passage of five
minimum ionising particles. Finally, at least one SPD hit is required among the SPD cells in front of the cluster.
 The software trigger requires a two-, three- or four-track
  secondary vertex with a high sum of the \pt of
  the tracks and a significant displacement from the primary $pp$
  interaction vertices~(PVs). At least one track should have $\pt >
  1.7\gevc$ and IP \chisq with respect to the
  primary interaction greater than 16. The IP \chisq is defined as the
  difference between the \chisq of the PV reconstructed with and
  without the considered track. A multivariate algorithm is used for
  the identification of secondary vertices consistent with the decay
  of a \bquark-hadron.

\par
The strategy of the analysis is to measure a ratio of branching fractions in which most of the 
potentially large systematic uncertainties cancel. The decay \BdToJPsieeKst is used as normalization mode, since it has the same final state as the \BdKstee decay and has a
well measured branching fraction~\cite{PDG2012,Aubert:2004rz}, approximately  300 times larger than $\BF(\BdKstee)$ in the
 \epem invariant mass range 30 to 1000 \mevcc.
Selection efficiencies are determined using data whenever possible, otherwise simulation is
used, with the events weighted to match the relevant distributions in data. 
The $pp$ collisions are generated using
\pythia~6.4~\cite{Sjostrand:2006za} with a specific \lhcb
configuration~\cite{LHCb-PROC-2010-056}.  Hadron decays 
are described by \evtgen~\cite{Lange:2001uf} in which final state
radiation is generated using \photos~\cite{Golonka:2005pn}. The
interaction of the generated particles with the detector and its
response are implemented using the \geant
toolkit~\cite{Allison:2006ve, *Agostinelli:2002hh} as described in
Ref.~\cite{LHCb-PROC-2011-006}.

\section{Selection and backgrounds}
\label{sec:SelB}
The candidate selection is divided into three steps: a loose selection, a multivariate algorithm to suppress the combinatorial background, and additional selection criteria to remove specific backgrounds.
\\
\par
Candidate \Kstarz mesons are reconstructed in the $\Kstarz \to K^+ \pi^-$ mode. The \pt of the charged $K$ ($\pi$) mesons must be larger than 400 (300) \mevc. Particle identification (PID) information is used to distinguish charged pions from kaons~\cite{LHCb-DP-2012-003}. The difference between the logarithms of the likelihoods of the kaon and pion hypotheses is required to be larger than 0 for kaons and smaller than 5 for pions; the combined efficiency of these cuts is 88\%. Candidates with a $K^+ \pi^-$ invariant mass within 130 
\mevcc of the nominal \Kstarz mass and a good quality vertex fit are retained for further analysis.
To remove  background from \BsToJPsieePhi and  \BsPhiee decays, where one of the kaons is misidentified as
a pion, the mass computed under the $K^+K^-$ hypothesis is required to be larger than 1040~\mevcc.  
\par
Bremsstrahlung radiation, if not accounted for,  would worsen the \Bz mass resolution. If the radiation occurs downstream of the dipole magnet the
momentum of the electron is correctly measured and the photon energy is deposited in the
same calorimeter cell as the electron. In contrast, if
photons are emitted upstream of the magnet, the measured electron momentum will be that after
photon emission, and the measured \Bz mass will be degraded. In general, these bremsstrahlung photons will deposit their energy in 
different calorimeter cells than the
electron.  In both cases, the ratio of the energy detected in the ECAL to the momentum measured by the tracking system, an important variable in identifying electrons, is unbiased. 
To improve the momentum reconstruction, a dedicated bremsstrahlung recovery
procedure is used, correcting the measured electron momentum by the bremsstrahlung
photon energy. As there is little material within the magnet, the bremsstrahlung photons are searched for among neutral clusters with an energy larger than 75 \mev in a well defined position given by the electron track extrapolation from before the magnet.
Oppositely-charged electron pairs with an electron \pt larger than 350 \mevc
and a good quality vertex are used to form \BdKstee and \BdToJPsieeKst candidates. 
The \epem invariant mass is required to be in the range 30 -- 1000 \mevcc or 2400 -- 3400 \mevcc for the two decay modes, respectively. 
Candidate \Kstarz mesons and \epem pairs are combined to form \Bd candidates which are required to have a good-quality vertex. For each \Bd candidate, the production vertex is assigned to be that with the smallest IP \chisq. 
The \Bd candidate is also required to have a direction that is consistent with coming from the PV as well as
a reconstructed decay point that is significantly separated from the PV. 

\begin{figure}[ht]
  \begin{center}
  \includegraphics[width=0.6\textwidth]{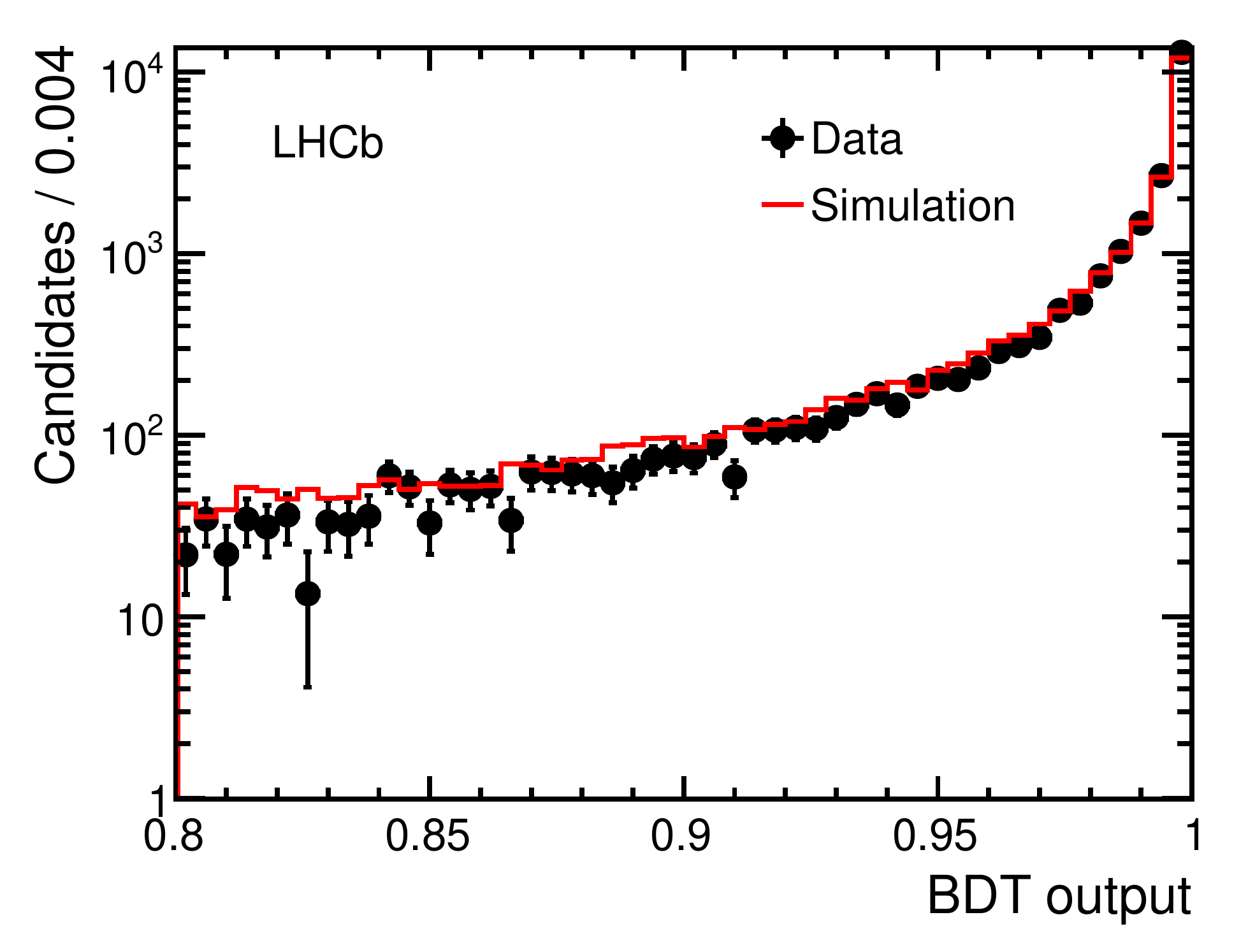}
 \vspace*{-1.0cm}
  \end{center}
  \caption{\small Output of the BDT for \BdToJPsieeKst data (points) and simulation (red line).}
  \label{fig:BDToutputsplot}
\end{figure}

In order to maximize the signal efficiency while still reducing the high level of combinatorial background, a multivariate analysis, based on a Boosted Decision Tree (BDT)~\cite{Breiman,*Roe} with the AdaBoost algorithm~\cite{AdaBoost}, is used.  The signal training sample is \BdKstee simulated data. The background training sample is taken from the upper sideband ($\mBd > 5600 \mevcc$) from half of the data sample.  The variables used in the BDT are the \pt , the IP and track \chisq of the final state particles; the \Kstarz candidate invariant mass, the vertex \chisq and flight distance \chisq (from the PV)  of the \Kstarz and \epem candidates; the \Bd \pt, its vertex \chisq, flight distance \chisq and IP \chisq, and the angle between the \Bd momentum direction and its direction of flight from the PV. 
A comparison of the BDT output for the data and the simulation for \BdToJPsieeKst decays is shown in Fig.~\ref{fig:BDToutputsplot}. The candidates for this test are reconstructed using a  \jpsi mass constraint and the background is statistically subtracted using the \sPlot\ technique~\cite{Pivk:2004ty} based on a fit to the \Bz invariant mass spectrum.  The agreement between data and simulation confirms a proper modelling of the relevant variables. The optimal cut value on the BDT response is chosen by considering the combinatorial background yield ($b$) on the \BdKstee invariant mass distribution outside the signal 
region\footnote{The signal region is defined as $\pm 300 \mevcc$ around the nominal \Bd mass.} and evaluating the signal yield ($s$) using the \BdKstee simulation assuming a visible  \BdKstee branching fraction of $2.7 \times 10^{-7}$. The quantity $s/\sqrt{s+b}$ serves  as an optimisation metric, for which the optimal BDT cut is 0.96. The signal efficiency of this cut is about 93\% while the background is reduced by two orders of magnitude.
\\
\par After applying the BDT selection, specific backgrounds from decays that have
the same visible final state particles as the \BdKstee signal remain. Since some of these backgrounds have
larger branching fractions, additional requirements are applied to the \BdKstee and \BdToJPsieeKst candidates. 

A large non-peaking background comes from the \decay{\Bd}{\Dm\ep\nu} decay, with  ${\Dm \ra \en { \overline \nu} \Kstarz}$.
The branching fraction for this channel is about five orders of magnitude larger than that of the signal. 
When the neutrinos have low energies, the signal selections are ineffective at rejecting this
background. Therefore, the $\Kstarz e^-$ invariant mass is required to be larger than 1900  \mevcc,
which is 97\% efficient on signal decays. Another important source of background comes from the \BdKstGam decay, where the photon converts into an \epem pair. In \lhcb, approximately 40\% of the photons convert before the calorimeter, and although only about 10\% are reconstructed as an \epem pair, the resulting mass of the \Bd candidate peaks in the signal region. This background is suppressed by a factor 23 after the selection cuts (including the 30\mevcc minimum requirement on the \epem invariant mass). The fact that signal \epem pairs are produced at the \Bd decay point, whereas conversion electrons
are produced in the VELO detector material, is exploited to further suppress this background. The difference in the $z$ coordinates, $\Delta z$, 
between the first VELO hit and the expected position of the first hit, assuming the electron was
produced at the \Kstarz vertex, should satisfy $| \Delta z| <30$ mm. In addition, we require that the calculated
uncertainty on the $z$-position of the \epem vertex be less than 30 mm, since a large uncertainty makes it difficult to determine if the \epem pair
originates from the same vertex as the \Kstarz meson, or from a point inside the detector material. 
These two additional  requirements reject about 2/3 of the remaining \BdKstGam background, while retaining about 90\% of the \BdKstee signal.
After applying these cuts, the \BdKstGam contamination under the \BdKstee signal peak  is estimated to be  $(10 \pm 3) \%$ of the expected signal yield. 

Other specific backgrounds have been studied using either simulated data or analytical calculations and include the decays $B\rightarrow K^{\ast}\eta , K^{\ast}\eta ' , K^{\ast}\pi^0$ and $\Lb \rightarrow \L^{\ast} \gamma$, where $\L^{\ast} $ represents a high mass resonance decaying into a proton and a charged kaon. The main source of background is found to be the  $B\rightarrow K^{\ast}\eta$ mode, followed by a Dalitz decay ($\eta \rightarrow \gamma \epem$). These events form an almost flat background in the mass range $4300 - 5250 \mevcc$. None of these backgrounds contribute significantly in the \Bd mass region, and therefore are not specifically modelled in the mass fits described later. 

More generally, partially reconstructed backgrounds arise from \B decays with one or more decay products in addition to a \Kstarz meson and an \epem pair. In the case of the \BdToJPsieeKst decay, there are two sources for these partially reconstructed events: those from the hadronic part, such as events with higher \Kstar resonances (partially reconstructed hadronic background), and those from the \jpsi part (partially reconstructed $J/\psi$ background), such as events coming from \psitwos decays. For the \BdKstee decay mode, only the partially reconstructed hadronic background has to be considered.

\section{Fitting procedure}
\label{sec:Fit}
Since the signal resolution,  type and rate of backgrounds depend on whether the hardware trigger was caused by a signal electron or by other
activity in the event, the data sample is divided into two mutually exclusive categories: events triggered 
by an extra particle $(e,\gamma,h,\mu)$ excluding the four final state particles (called HWTIS, since they are triggered independently of the signal) and events for which one of the electrons from the \Bz decay satisfies the hardware electron trigger (HWElectron). Events satisfying both requirements (20\%) are assigned to the HWTIS category. 
The numbers of reconstructed signal candidates are determined from unbinned maximum likelihood fits to their mass distributions separately for each trigger category. The mass distribution of each category is fitted to a sum of probability density functions (PDFs) modelling the different components.
\begin{enumerate}
\item The signal is described by the sum of two Crystal Ball functions~\cite{Skwarnicki:1986xj} (CB) sharing all their parameters but with different widths.
\item The combinatorial background is described by an exponential function.
\item The shapes of the partially reconstructed hadronic and \jpsi backgrounds are described by non-parametric PDFs~\cite{Cranmer:2000du} determined from fully simulated events. 
\end{enumerate}
The signal shape parameters are fixed to the values obtained from simulation, unless otherwise specified. 

\begin{figure}[htpb]
\begin{center}
\includegraphics[width=0.495\textwidth]{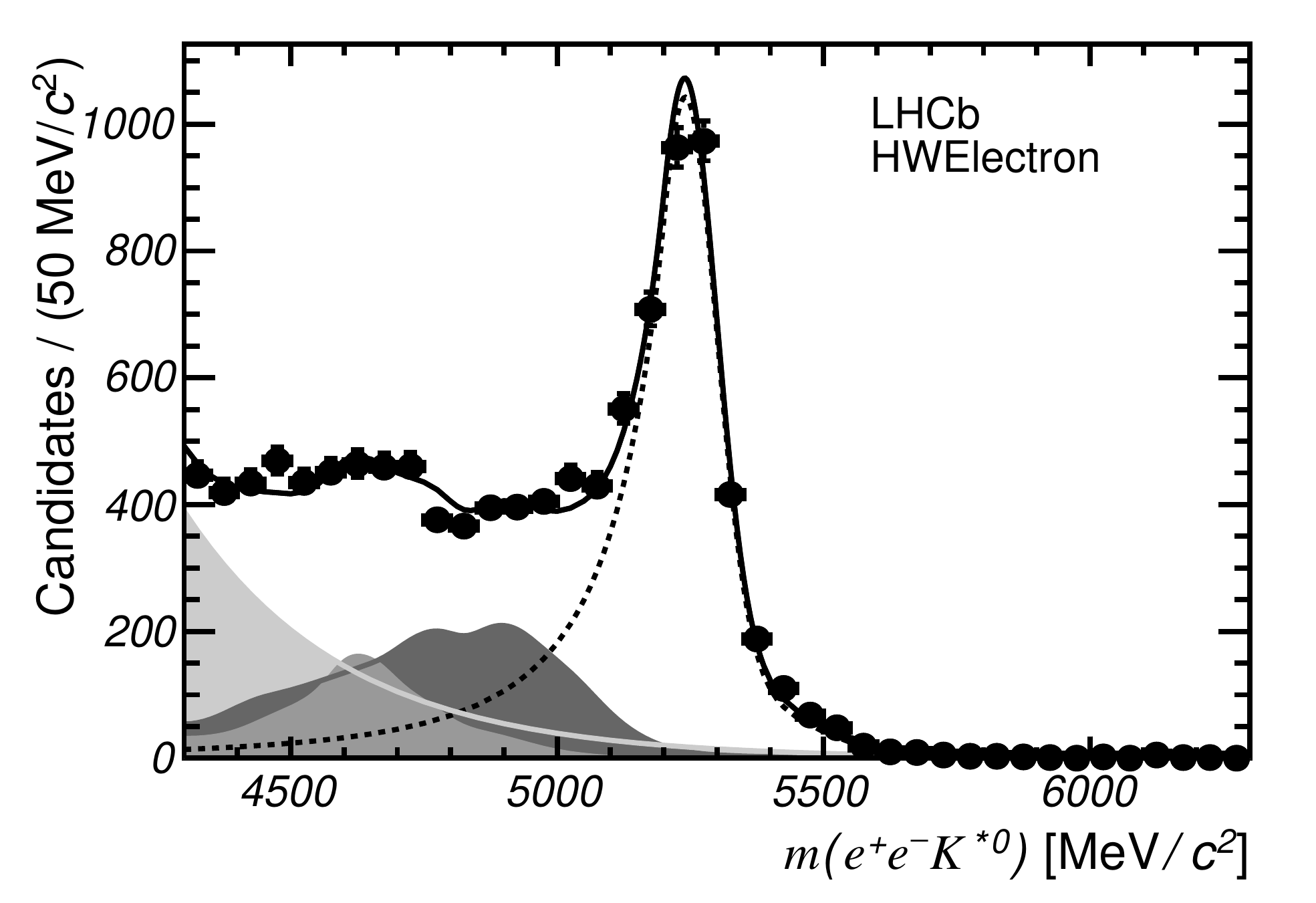} 
\includegraphics[width=0.495\textwidth]{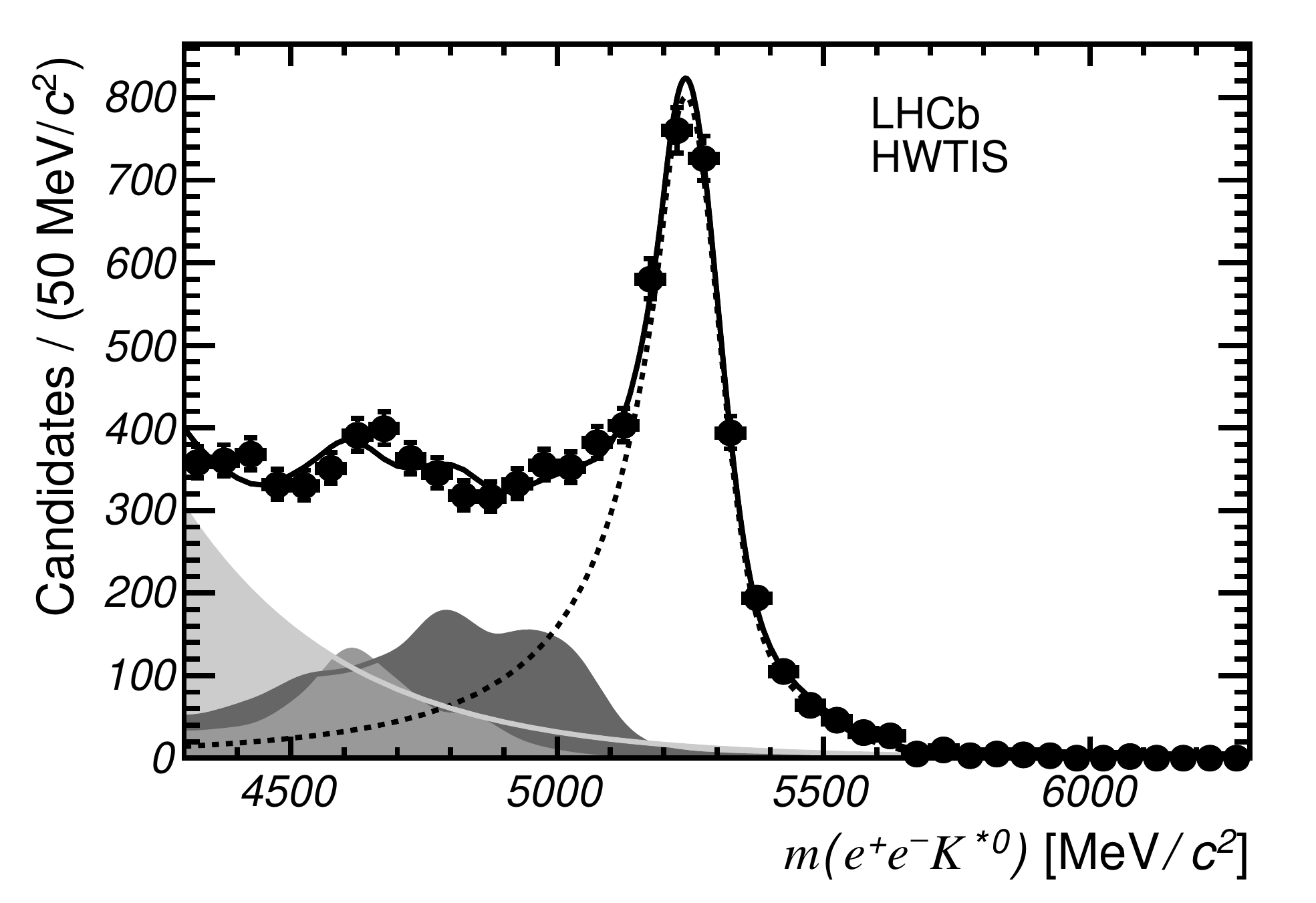} 
\end{center}
\caption{\small Invariant mass distributions for the  \BdToJPsieeKst decay mode for the (left) HWElectron and (right) HWTIS trigger categories.
The dashed line is the signal PDF, the light grey area corresponds to the combinatorial background, the medium grey area is the partially reconstructed hadronic background and the dark grey area is the partially reconstructed \jpsi background component.}Ê
\label{fig:JpsiKstar} 
\end{figure}

There are seven free parameters for the \BdToJPsieeKst fit for each trigger category. These include the peak value of the \Bd candidate mass, a scaling factor applied to the widths of the CB functions to take into account small differences between simulation and data, and the exponent of the combinatorial background. The remaining four free parameters are the yields for each fit component. 
The invariant mass distributions together with the PDFs resulting from the fit are shown in Fig.~\ref{fig:JpsiKstar}. The number of signal events in each category is summarized in 
Table~\ref{tab:yields}. 

\begin{figure}[htpb]
\begin{center}
\includegraphics[width=0.495\textwidth]{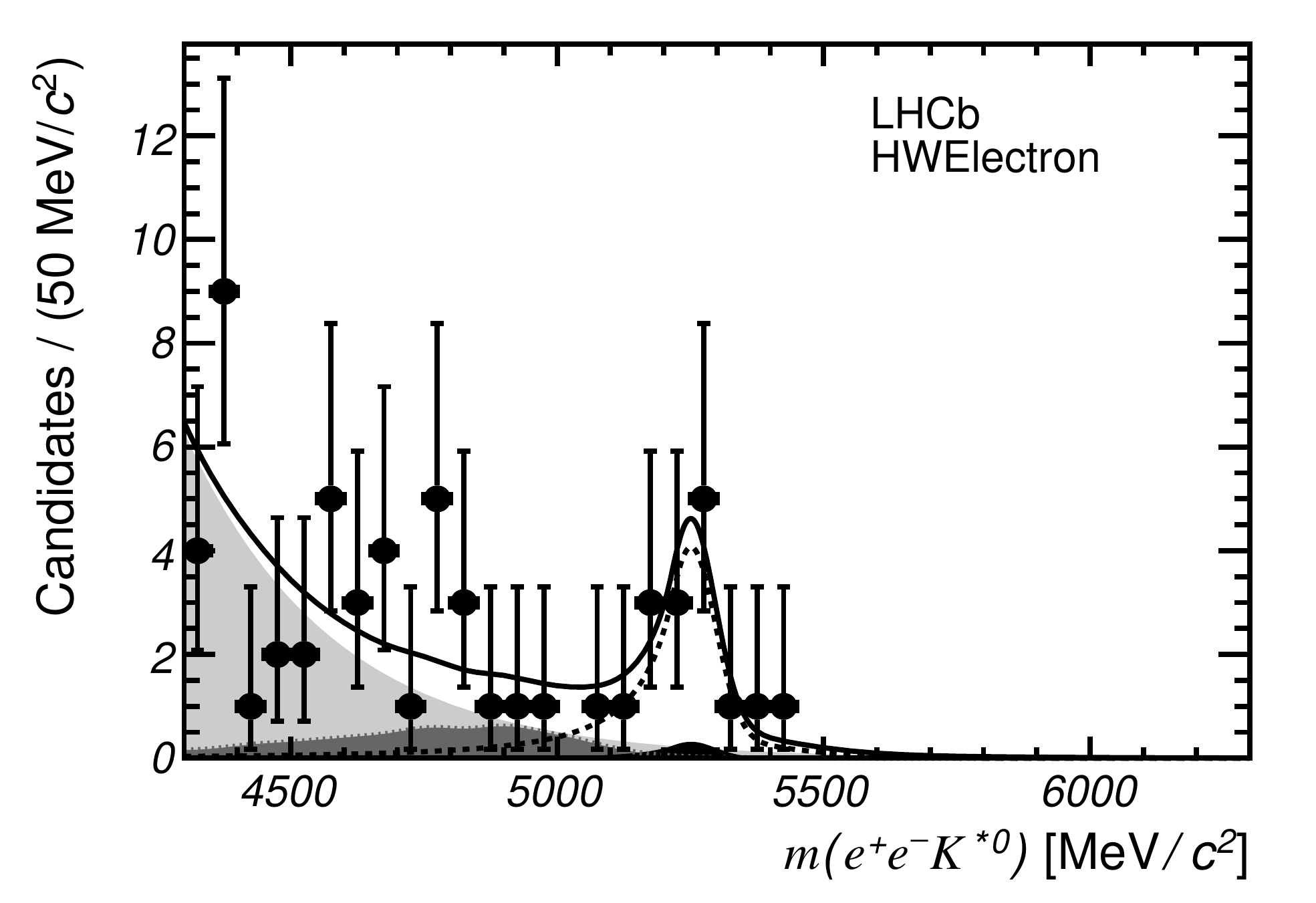} 
\includegraphics[width=0.495\textwidth]{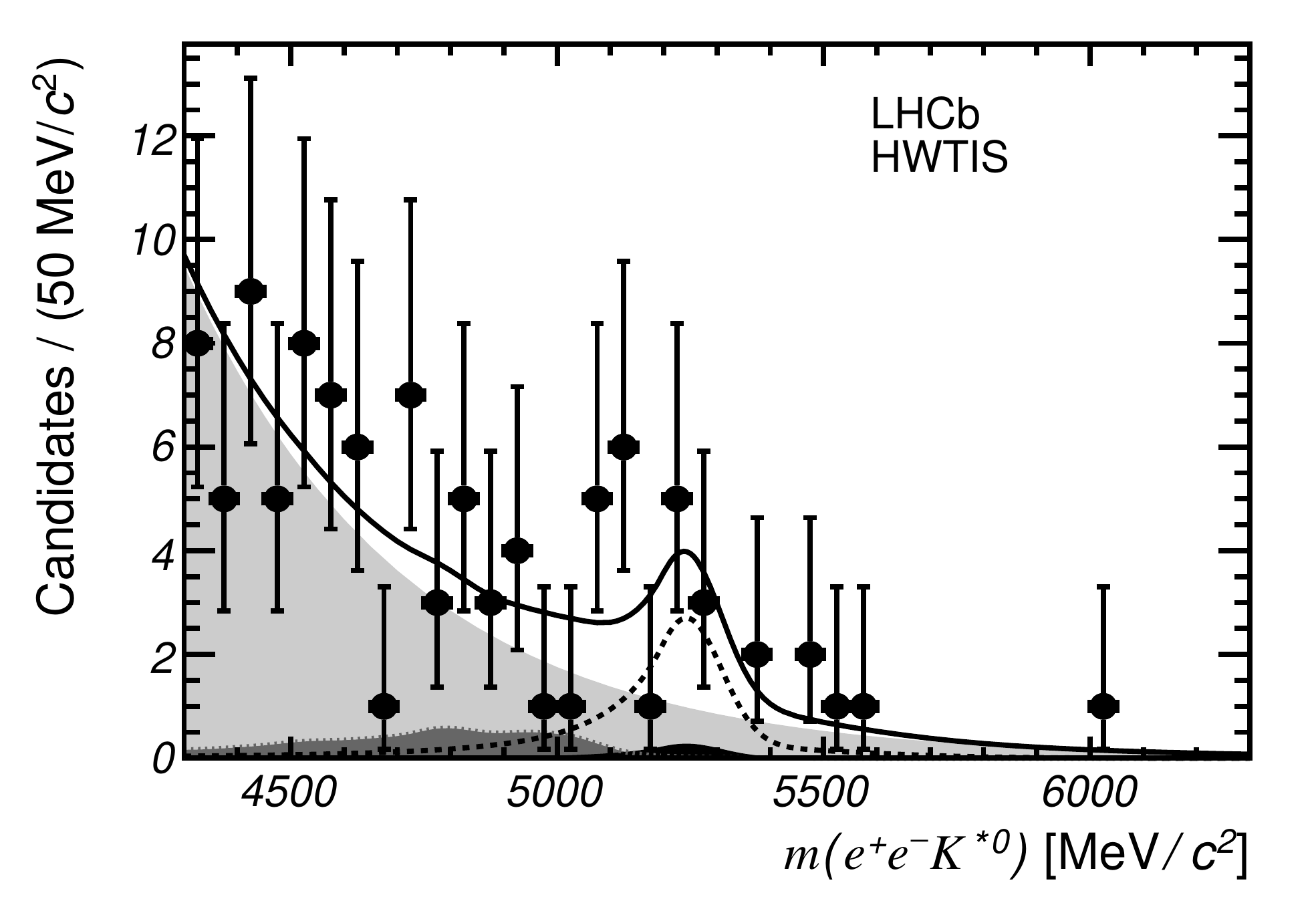} 
\end{center}
\caption{\small Invariant mass distributions for the  \BdKstee decay mode for the (left) HWElectron and (right) HWTIS trigger categories.
The dashed line is the signal PDF, the light grey area corresponds to the combinatorial background, the medium grey area is the partially reconstructed hadronic background and the black area is the \BdKstGam component.}Ê
\label{fig:eeKstar} 
\end{figure}

A fit  to the \BdKstee candidates is then performed, with several parameters fixed to the values found from the \BdToJPsieeKst fit. These fixed parameters are the scaling factor applied to the widths of the CB functions, the peak value of the \Bd candidate mass and the ratio of the partially reconstructed hadronic background to the signal yield. The \BdKstGam yield is fixed in the \BdKstee mass fit
using the fitted \BdToJPsieeKst signal yield, the ratio of
efficiencies of the \BdKstGam and \BdToJPsieeKst modes, and the ratio of
branching fractions $\BR(\BdKstGam)/\BR(\BdToJPsieeKst)$. Hence there are three free parameters for the \BdKstee fit for each trigger category: the exponent and yield of the combinatorial background and the signal yield. The invariant mass distributions together with the PDFs resulting from the fit are shown in Fig.~\ref{fig:eeKstar}. The signal yield in each trigger category is summarized in  Table~\ref{tab:yields}.  The probability of the background fluctuating to obtain the observed signal corresponds to 4.1 standard deviations for the HWElectron category and 2.4  standard deviations for the HWTIS category, as determined from the change in the value of twice the natural logarithm of the likelihood of the fit with and without signal. Combining the two results, the statistical significance of the signal
corresponds to 4.8 standard deviations.

 \begin{table}[htpb]
 \caption{\small Signal yields with their statistical uncertainties.}
\begin{center}
\renewcommand{\arraystretch}{1.2}
\begin{tabular}{lcc}
Trigger category &  \BdToJPsieeKst &   \BdKstee \\
  \hline
    HWElectron 	& $5082 \pm 104$ &   $15.0\,^{+5.1}_{-4.5}$ \\
    HWTIS	 	& $4305 \pm 101$ &    $14.1\, ^{+7.0}_{-6.3}$ \\
  \end{tabular}
\end{center}
\label{tab:yields}
\end{table}

\section{Results}
\label{sec:BR}
The \BdKstee branching fraction is calculated in each trigger category using the measured signal yields and the ratio of efficiencies
\begin{eqnarray}
\BF(\BdKstee)^{30-1000 \mevcc} =& \frac{N(\BdKstee)}{N(\BdToJPsieeKst)} 
 \times r_{\rm sel} \times r_{\rm PID} \times r_{\rm HW} \\ \nonumber
& \times \BF(\BdToJPsiKst) \times \BF(\JPsiToee),
\label{eq:br}
\end{eqnarray}
where the ratio of efficiencies is sub-divided into the contributions arising from the selection requirements (including acceptance effects, but excluding PID), $r_{\rm sel}$, the PID requirements $r_{\rm PID}$ and the trigger requirements $r_{\rm HW}$. The values of $r_{\rm sel} $ are determined using simulated data, while $r_{\rm PID}$ and $r_{\rm HW}$ are obtained directly from calibration data samples: \JPsiToee and $\Dz \ra K^-\pi^+$ from $\Dstarp$ decays for  $r_{\rm PID}$ and \BdToJPsieeKst decays for $r_{\rm HW}$. The values are summarized in Table~\ref{tab:eff}. The only ratio that is inconsistent with unity is the hardware trigger efficiency due to the different mean electron \pt for the \BdKstee and \BdToJPsieeKst decays. 
\par
The branching fraction for the \BdToJPsiKst decay mode is taken from Ref.~\cite{Aubert:2004rz} and a correction factor of 1.02 has been applied to take into account the difference  in the $K \pi$ invariant mass range used, and therefore the different S-wave contributions.

\begin{table}[tpb]
\caption{\small Ratios of efficiencies used for the measurement of the \BdKstee branching fraction. The ratio $r_{\rm HW}$ for the HWTIS trigger category is assumed to be equal to unity. The uncertainties are the total ones and are discussed in Sec.~\ref{sec:syst}.}
\begin{center}\begin{tabular}{lcc}
				& HWElectron category 			& HWTIS 	category		\\
\hline
$r_{\rm sel}$ 		& $1.03 \pm 0.02$ 		& $1.03 \pm 0.02$ \\  
$r_{\rm PID}$ 		& $1.01 \pm 0.02$& $1.03 \pm 0.02$\\  
$ r_{\rm HW}$		 & $1.35 \pm 0.03$ & 1 \\ 
 \end{tabular}
\end{center}
\label{tab:eff}
\end{table}

The \BdKstee branching fraction, for each trigger category, is measured to be
\begin{eqnarray}
\BF(\BdKstee)^{30-1000 \mevcc}_{\text{HWElectron}} &=& (3.3\,^{+1.1}_{-1.0}) \times 10^{-7} \nonumber \\ 
\BF(\BdKstee)^{30-1000 \mevcc}_{\text{HWTIS}} &=& (2.8\,^{+1.4}_{-1.2}) \times 10^{-7}, \nonumber
\end{eqnarray}
where the uncertainties are statistical only.

\section{Systematic uncertainties \label{sec:syst}}
Several sources of systematic uncertainty are considered, affecting either the determination of the number of signal events or the computation of the efficiencies. 
They are summarized in Table~\ref{tab:syst}. 

The ratio of trigger efficiencies is determined using a \BdToJPsieeKst  calibration sample from data, which is reweighted using the \pt of the triggering electron in order to model properly the kinematical properties of the two decays. The uncertainties due to the limited size of the calibration samples are propagated to get the related systematic uncertainty shown in Table~\ref{tab:eff}. 

The PID calibration introduces a systematic uncertainty on the calculated PID efficiencies as given in Table~\ref{tab:eff}. For the kaon and pion candidates this systematic uncertainty is estimated by comparing, in simulated events, the results obtained using a \Dstarp calibration sample to the true simulated PID performance.
For the \epem candidates, the systematic uncertainty is assessed ignoring the \pt dependence of the electron identification. The resulting effect is limited by the fact that the kinematic differences between the \BdToJPsieeKst  and the \BdKstee  decays are small once the full selection chain is applied. 
\par
The fit procedure is validated with pseudo-experiments. Samples are generated with different fractions or shapes for the partially reconstructed hadronic background, or different values for the fixed signal parameters and are then fitted with the standard PDFs. The corresponding systematic uncertainty is estimated from the bias in the results obtained by performing the fits described above. The resulting deviations from zero of each variation are added in quadrature to get the total systematic uncertainty due to the fitting procedure. The parameters of the signal shape are varied within their statistical uncertainties as obtained from the \BdToJPsieeKst fit. An alternate signal shape, obtained by studying \BdToJPsieeKst signal decays in data
both with and without a \jpsi mass constraint is also tried; the difference in the yields from that obtained using the nominal signal shape is taken as an additional source of uncertainty. The ratio of the partially reconstructed hadronic background to the signal yield is assumed to be identical to that determined from the  \BdToJPsieeKst fit. The systematic uncertainty linked to this hypothesis is  evaluated by varying the ratio by $\pm 50\%$. The fraction of partially reconstructed hadronic background thus determined is in agreement within errors with the one found in \BdKstGam decays~\cite{Aaij:2012ita}. The shape of the partially reconstructed background used in the \BdToJPsieeKst and the  \BdKstee fits are the same. The related systematic uncertainty has been evaluated  using an alternative shape obtained from charmless  \bquark-hadron decays.  
The  \BdKstGam contamination in the \BdKstee  signal sample is $1.2 \pm 0.4$ and $1.5 \pm 0.5$ events for the HWElectron and HWTIS signal
samples, respectively. Combining the systematic uncertainties in quadrature, the branching fractions are found to be
\begin{eqnarray}
\BF(\BdKstee)^{30-1000 \mevcc}_{\rm HWElectron}&=& (3.3\,^{+1.1\mbox{ } +0.2}_{-1.0\mbox{ }-0.3 } \pm 0.2) \times 10^{-7} \nonumber \\ 
\BF(\BdKstee)^{30-1000 \mevcc}_{\rm HWTIS}&=& (2.8\,^{+1.4\mbox{ } +0.2}_{-1.2\mbox{ }-0.3 } \pm 0.2)\times 10^{-7}, \nonumber
\end{eqnarray}
where the first error is statistical, the second systematic, and the third comes from the uncertainties on the \BdToJPsiKst and \JPsiToee  branching fractions~\cite{PDG2012,Aubert:2004rz}. 
The branching ratios are combined assuming all the systematic uncertainties to be fully correlated between the two trigger categories except those related to the size of the simulation samples. The combined branching ratio is found to be
\begin{equation}
\BF(\BdKstee)^{30-1000 \mevcc}= (3.1\,^{+0.9\mbox{ } +0.2}_{-0.8\mbox{ }-0.3 }  \pm 0.2)\times 10^{-7}. \nonumber
\label{eq:br41}
\end{equation}
   
\begin{table}[tpb]
\caption{\small Absolute systematic uncertainties on the \BdKstee branching ratio (in $10^{-7}$) .}
\begin{center}
\renewcommand{\arraystretch}{1.2}
\begin{tabular}{lcc}            
Source   					& HWElectron category 		& HWTIS category 		\\ 
\hline
Simulation sample statistics 				& 	0.06			& 0.05 \\
Trigger efficiency 			& 	0.07  		&\--\\       
PID efficiency     			& 	0.08			&  0.10  \\
Fit procedure     			&   $^{+0.09}_{-0.22}$    & $^{+0.07}_{-0.23}$\\                                       
 \BdKstGam   contamination 	& 0.08 & 0.08\\         
                \hline                                 
 Total  					& $^{+0.17}_{-0.26}$ & $^{+0.16}_{-0.27}$\\    
                  \end{tabular}
\end{center}
\label{tab:syst}
\end{table}

\section{Summary}
Using $pp$ collision data corresponding to an integrated luminosity of 1.0~\invfb, collected by the \lhcb experiment in 2011 at a centre-of-mass energy of 7\tev, a sample of approximately 30 \BdKstee events, in the dilepton mass range 30 to 1000\mevcc, has been observed.  The probability of the background to fluctuate upward to form the signal corresponds to 4.6 standard deviations including systematic uncertainties.
The \BdToJPsieeKst decay mode is utilized as a normalization channel, and the branching fraction \BF(\BdKstee) is measured to be
$$
\BF(\BdKstee)^{30-1000 \mevcc}= (3.1\,^{+0.9\mbox{ } +0.2}_{-0.8\mbox{ }-0.3 }  \pm 0.2)\times 10^{-7}.
$$

This result can be compared to theoretical predictions. A simplified formula suggested in Ref.~\cite{Grossman:2000rk} takes into account only the photon diagrams of  Fig.~\ref{fig:feynman}. When evaluated
in the 30 to 1000\mevcc \epem invariant mass interval using \BR(\BdKstGam)~\cite{Aubert:2009ak,Nakao:2004th,Coan:1999kh}, it predicts a \BdKstee
branching fraction of $2.35 \times 10^{-7}$. A full calculation has been recently performed~\cite{Jager:2012uw} and the numerical result for the 
 \epem invariant mass interval of interest is $ (2.43^{+0.66}_{-0.47} )\times 10^{-7} $. The consistency between the two values 
reflects the photon pole dominance. The result presented here is in good agreement with both predictions. 
\par
Using the full \lhcb data sample obtained in 2011 -- 2012 it will be possible 
to do an angular analysis. The measurement of the \AT2 parameter~\cite{Kruger:2005ep} thus
obtained, is sensitive to the existence of right handed currents in the virtual
loops in diagrams similar to those of Fig. \ref{fig:feynman}. For this purpose, the analysis of
the \BdKstee decay is complementary to that of the \BdToKstmm mode. Indeed, it is predominantly sensitive to a modification of \C{7} 
(the so-called \Cp{7} terms) while, because of the higher \qsq in the decay, the
\BdToKstmm \AT2 parameter has a larger possible contribution from the \Cp{9}  terms~\cite{Becirevic:2011bp}.

\section*{Acknowledgements}

\noindent We express our gratitude to our colleagues in the CERN
accelerator departments for the excellent performance of the LHC. We
thank the technical and administrative staff at the LHCb
institutes. We acknowledge support from CERN and from the national
agencies: CAPES, CNPq, FAPERJ and FINEP (Brazil); NSFC (China);
CNRS/IN2P3 and Region Auvergne (France); BMBF, DFG, HGF and MPG
(Germany); SFI (Ireland); INFN (Italy); FOM and NWO (The Netherlands);
SCSR (Poland); ANCS/IFA (Romania); MinES, Rosatom, RFBR and NRC
``Kurchatov Institute'' (Russia); MinECo, XuntaGal and GENCAT (Spain);
SNSF and SER (Switzerland); NAS Ukraine (Ukraine); STFC (United
Kingdom); NSF (USA). We also acknowledge the support received from the
ERC under FP7. The Tier1 computing centres are supported by IN2P3
(France), KIT and BMBF (Germany), INFN (Italy), NWO and SURF (The
Netherlands), PIC (Spain), GridPP (United Kingdom). We are thankful
for the computing resources put at our disposal by Yandex LLC
(Russia), as well as to the communities behind the multiple open
source software packages that we depend on.

\addcontentsline{toc}{section}{References}
\bibliographystyle{LHCb}
\bibliography{main,LHCb-PAPER,LHCb-CONF,LHCb-DP}

\end{document}